\documentclass{eas}
\usepackage{graphicx}
\begin{document}

\title{Four open questions in massive star evolution} 
 \runningtitle{four open questions}
\author{Georges Meynet}\address{Geneva Observatory, University of Geneva, CH-1290 Versoix Switzerland}
\author{Patrick Eggenberger}\sameaddress{1}
\author{Sylvia Ekstr\"om}\sameaddress{1}
\author{Cyril Georgy}\address{Astrophysics group, Lennard-Jones Laboratories, EPSAM, Keele University, Staffordshire, ST5 5BG}
\author{Jos\'e Groh}\sameaddress{1}
\author{Andr\'e Maeder}\sameaddress{1}
\author{Hid\'eyuki Saio}\address{Astronomical Institute, Graduate School of Science, Tohoku University, Sendai, Japan}
\author{Takashi Moriya}\address{Kavli Institute for the Physics and Mathematics of the Universe (WPI), Todai Institutes for Advanced 
Study, University of Tokyo, Kashiwanoha 5-1-5, Kashiwa, Chiba 277-8583, Japan}
%
%
\begin{abstract}
We discuss four questions dealing with massive star evolution. The first one is about the origin of slowly rotating, non-evolved, nitrogen rich stars. We propose that these stars 
may originate from initially fast rotating stars whose surface has been braked down. 
The second question is about the evolutionary status of $\alpha$-Cygni variables. According to their
pulsation properties, these stars should be post red supergiant stars. However, some stars at least present surface abundances indicating that they should be pre red supergiant stars. How to reconcile these two contradictory requirements? 
The third one concerns the various supernova types which are the end point of the evolution of stars with initial masses between 18 and 30 M$_\odot$, {\it i.e.}
the most massive stars which go through a red supergiant phase during their lifetime. Do they produce types IIP, IIL,  IIn,   IIb or  Ib  supernovae or do they end without producing any SN event?
Finally, we shall discuss reasons why so few progenitors of type Ibc supernovae have yet been detected? 
\end{abstract}
\maketitle
\section{Puzzle 1: What is the origin of the N-rich, non-evolved, slowly rotating stars?}

The authors would like to present their greetings to Sylvie Vauclair whose scientific activity has
produced so wonderful achievements in the field of stellar physics.
The work done by Sylvie Vauclair very well reflects
the qualities that Jean Rostand (1894-1977), a french biologist, writer and to some extent philosopher, see to be in the word researcher: {\it ``Beau mot que celui de chercheur, et si pr\'ef\'erable \`a celui de savant ! Il exprime la saine attitude de l'esprit devant la v\'erit\'e : le manque plus que l'avoir, le d\'esir plus que la possession, l'app\'etit plus que la sati\'et\'e\footnote{Beautiful word the one of  researcher, and so preferable to scholar! It expresses the healthy attitude of the mind in front of the truth: the lack and the desire more than the ownership, the appetite more than the satisfaction. {\it Tentative translation of G.M.}}}.

In the recent years, many efforts have been made to study the impact of rotation on massive star evolution (see e.g. Langer \cite{2012ARA&A..50..107L}, Maeder \& Meynet \cite{2012RvMP...84...25M}, Chieffi \& Limongi \cite{2013ApJ...764...21C} and references therein). One of the main effects of rotation is to 
transport some elements abundant in the core to the surface and inversely some other elements, abundant in the envelope, into the convective core.
This rotational mixing process has many consequences for the evolutionary tracks in the HR diagram, the lifetimes, the massive star populations, the nucleosynthesis, the properties
of the final stellar collapse and of the stellar remnants. The understanding of the physics of rotation is not only 
important for the modeling of single stars but also of close binaries.

 One way of checking the physics included in these models is to observe surface
 abundances and surface velocities of B-type stars and to see whether the observations agree with the predictions of the theoretical models (see e.g. Hunter {\em et al.} \cite{2009A&A...496..841H}; Brott {\em et al.} \cite{2011A&A...530A.116B}, Przybilla {\em et al.} \cite{2010A&A...517A..38P}). For making relevant comparisons, the observed sample must be composed of 
stars with different surface velocities but with all other characteristics being as similar as possible  ({\it i.e.} same initial mass, initial composition, age, same magnetic field if any, no close companion).
When such precautions are taken, general good agreement is obtained between theory and observations (Maeder {\em et al.} \cite{2009CoAst.158...72M}). 

There is however a small group of stars  which does not appear to follow the general trend predicted by the rotating stellar models. 
They present three properties which are difficult to reconcile: they show 1.- low $\upsilon\sin i$ values, where $\upsilon$ is the surface equatorial velocity and $i$, the angle between the direction along the line of sight and the rotational axis, 2.- strong nitrogen enrichment at the surface and 3.- they are non-evolved stars. This last condition implies that the enrichment should have occurred fast. The problem is then how to obtain strong N-rich stars, in a short time with a low surface velocity?  Are all these stars fast rotators seen pole on? This may be the case for a small fraction of them\footnote{In Hunter {\em et al.} \cite{2009A&A...496..841H}, 12 stars show properties 1 to 3 in a sample containing 135 early B-type stars in the LMC.} but it would be unreasonable to invoke this explanation for the whole observed sample.

Another possibility is the following: the large nitrogen enrichment results from a strong internal differential rotation produced by a braking mechanism acting on the surface layers. The braking can be induced by a surface magnetic field strong enough to couple with the stellar wind. A nice example of this process is provided by the star $\sigma$ Ori E, whose rotation period  (1.19 days) increases of 77 milliseconds per year (Townsend {\em et al.} \cite{2010ApJ...714L.318T}). Very interestingly, this observed increase of the period
is well reproduced by the model of magnetic wind braking proposed by Ud Doula {\em et al.} ( \cite{2009MNRAS.392.1022U}).

\begin{figure}
\includegraphics[width=2.2in,height=2.2in]{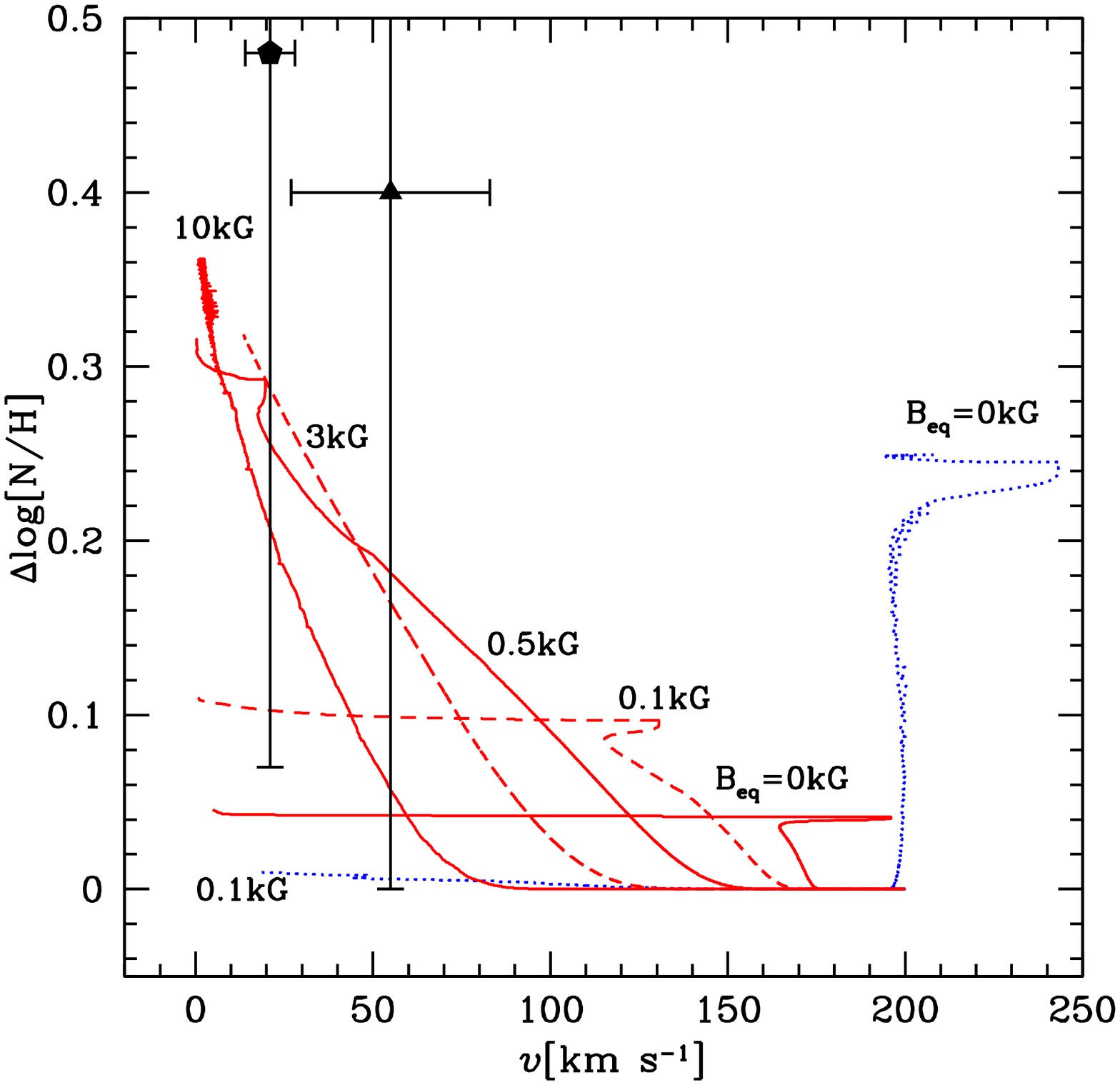}
\hfill
\includegraphics[width=2.2in,height=2.2in]{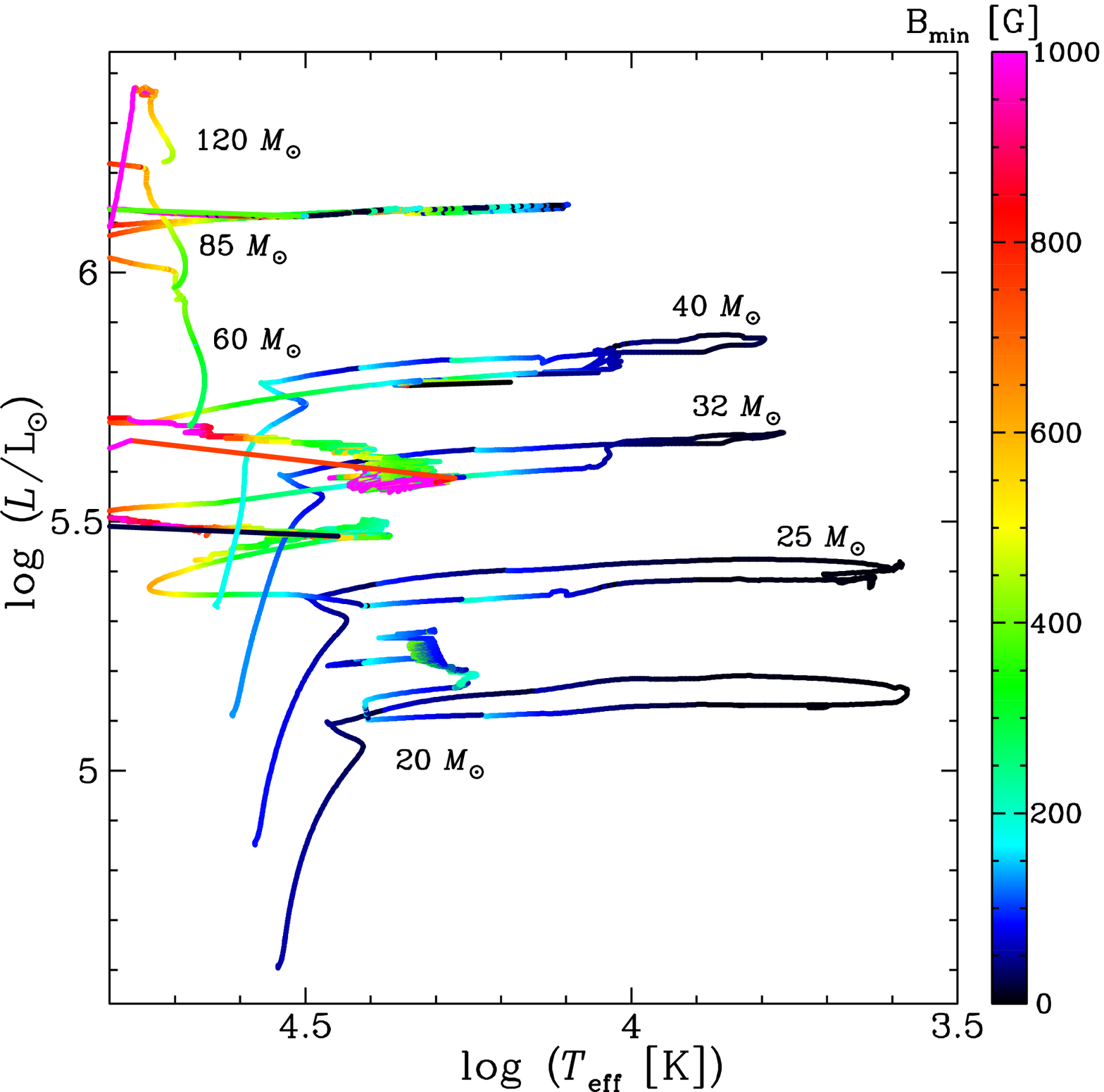}
\caption{{\it Left panel}: Variation of the surface N abundance as a function of the surface rotational velocity for 10 M$_\odot$ stellar models at solar metallicity with an initial velocity on the ZAMS equal to 200 km s$^{-1}$ (Meynet {\em et al.} \cite{2011A&A...525L..11M}). The continuous and dashed lines correspond to models computed with shear and meridional currents, the dotted lines correspond to models computed with solid body rotation and meridional currents. The surface magnetic fields used for the braking are indicated. The pentagon and the triangle show observed values for HD 16582 ($\xi$ Cas) and HD 3360 (see references in Meynet {\em et al.} \cite{2011A&A...525L..11M}).
{\it Right panel}: Evolutionary tracks in the HR diagram. The colors along the tracks indicate the minimum magnetic field (assumed to be aligned with the rotation axis and bipolar) able to couple the stellar wind with the stellar surface and thus to exert a torque. The minimum value is obtained by imposing that the parameter $\eta$ defined in Ud-Doula \& Owocki (\cite{2009MNRAS.392.1022U}) is equal to one. 
Figure taken from Georgy  {\em et al.} \cite{2012A&A...542A..29G}.
}
\label{fig1}
\end{figure}

To know, whether such a mechanism could indeed be interesting to provide an explanation for these stars, we have
computed 10 M$_\odot$ rotating models for solar metallicity, accounting for the magnetic braking of the surface according to the recipe suggested by Ud-Doula {\em et al.} (\cite{2009MNRAS.392.1022U}; see Meynet {\em et al.} \cite{2011A&A...525L..11M}). In Fig.~\ref{fig1}, the evolution
of the surface nitrogen abundance is plotted as a function of the surface equatorial velocity. Two cases are shown. In the first one
(look at the continuous and dashed lines), the interior of the star is rotating differentially and the mixing is mainly driven by the shear, while in the second case (dotted lines), the interior is rotating as a solid body and the mixing is driven by meridional currents.

In differentially rotating star, we note that the magnetic braking can  produce main-sequence stars which are slowly rotating and nitrogen enriched. At the moment very few of these stars are observed at solar metallicity, two of them are plotted in Fig.~\ref{fig1} (left panel)\footnote{Many more have been observed in the LMC
sample of the VLT Flames survey, see Hunter {\em et al.} \cite{2009A&A...496..841H}, but the models computed here are for solar metallicity, so we restrain the comparisons to this metallicity.}. The error bars are quite large and the comparison is therefore not very constraining. But we can see that in principle the mechanism could indeed work. It is interesting to mention that
one of the star plotted (the pentagon) has a detected surface magnetic field of about 335$+120-65$G (Neiner {\em et al.}, \cite{2003A&A...406.1019N}).

In a solid body rotating star, on the other hand, the braking mechanism damps the source of the chemical mixing (meridional circulation)
before it can produce any effect. Thus, only slow rotators would be expected in that case with no nitrogen enrichment. These stars would be similar to truely initially slow rotators, except that they can present a very slight depletion of fragile elements like Boron (Meynet {\em et al.} \cite{2011A&A...525L..11M}).

In case the present explanation would be the correct one, it would mean that non-evolved stars with high surface N content, low $\upsilon\sin i$ stars would be magnetic stars. Unless the magnetic field would have decayed or be no long sustained by a dynamo process due to the slowing down of the star, one would expect
to observe significant surface magnetic field  for these stars. Another interesting point is that, these stars should not be solid body rotating star, at least during the braking period because otherwise, no nitrogen enrichment would be expected.

The fact that only a small fraction of stars belong to these group 2 stars, indicate that most of the massive stars should not have a
very strong surface magnetic fields. From the left panel of Fig.~\ref{fig1}, we see that a minimum surface magnetic field of about 500 G is needed to produce group 2 stars.
In Fig.~\ref{fig1} (right panel), the minimum surface magnetic field for obtaining a
magnetic braking is indicated for various initial mass stars (Georgy  {\em et al.} \cite{2012A&A...542A..29G}). 
We see that for stars with masses below about 30 M$_\odot$, already a surface magnetic field above 100 G would already be sufficient to exert some coupling. It means that
if a large majority of stars in this mass range would have a surface magnetic field of the order of 100 G then most of the stars would be very slow rotators with no or
with strong N-enrichments (depending on internal rotation profile) which is not observed. This does appear quite in line with the results of the recent MIMES (Magnetism in Massive Stars, see e.g. Grunhut {\em et al.} \cite{2012ASPC..465...42G}) indicating that 
only 6-7\% of OB stars show surface magnetic fields with intensities above 0.1-2.0 kG (the detectability depends on the spectral type). 

\section{Puzzle 2: What is the origin of the pulsating properties of alpha Cygni stars? }

\begin{figure}
\includegraphics[width=2.2in,height=2.2in]{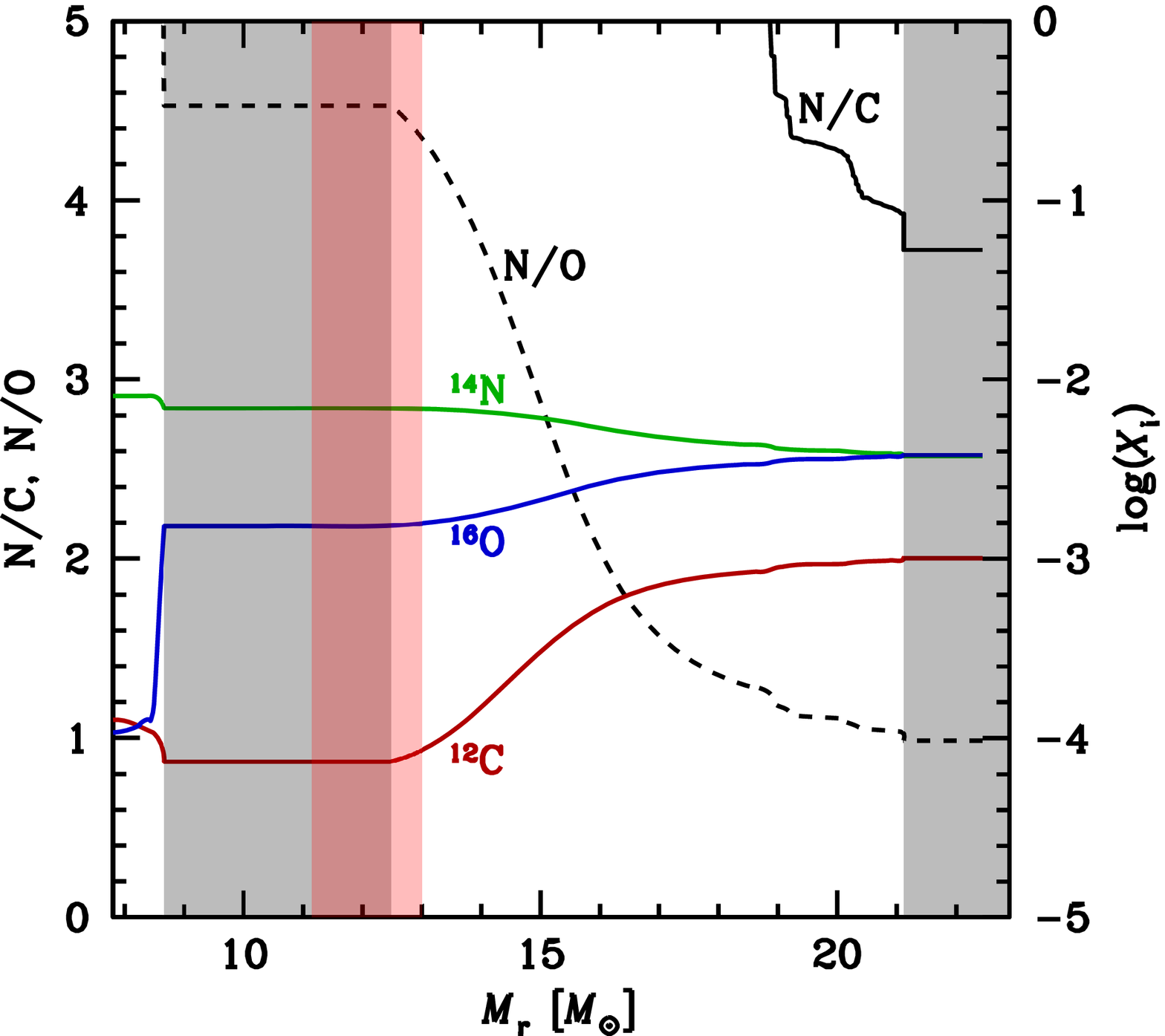}
\hfill
\includegraphics[width=2.2in,height=2.2in]{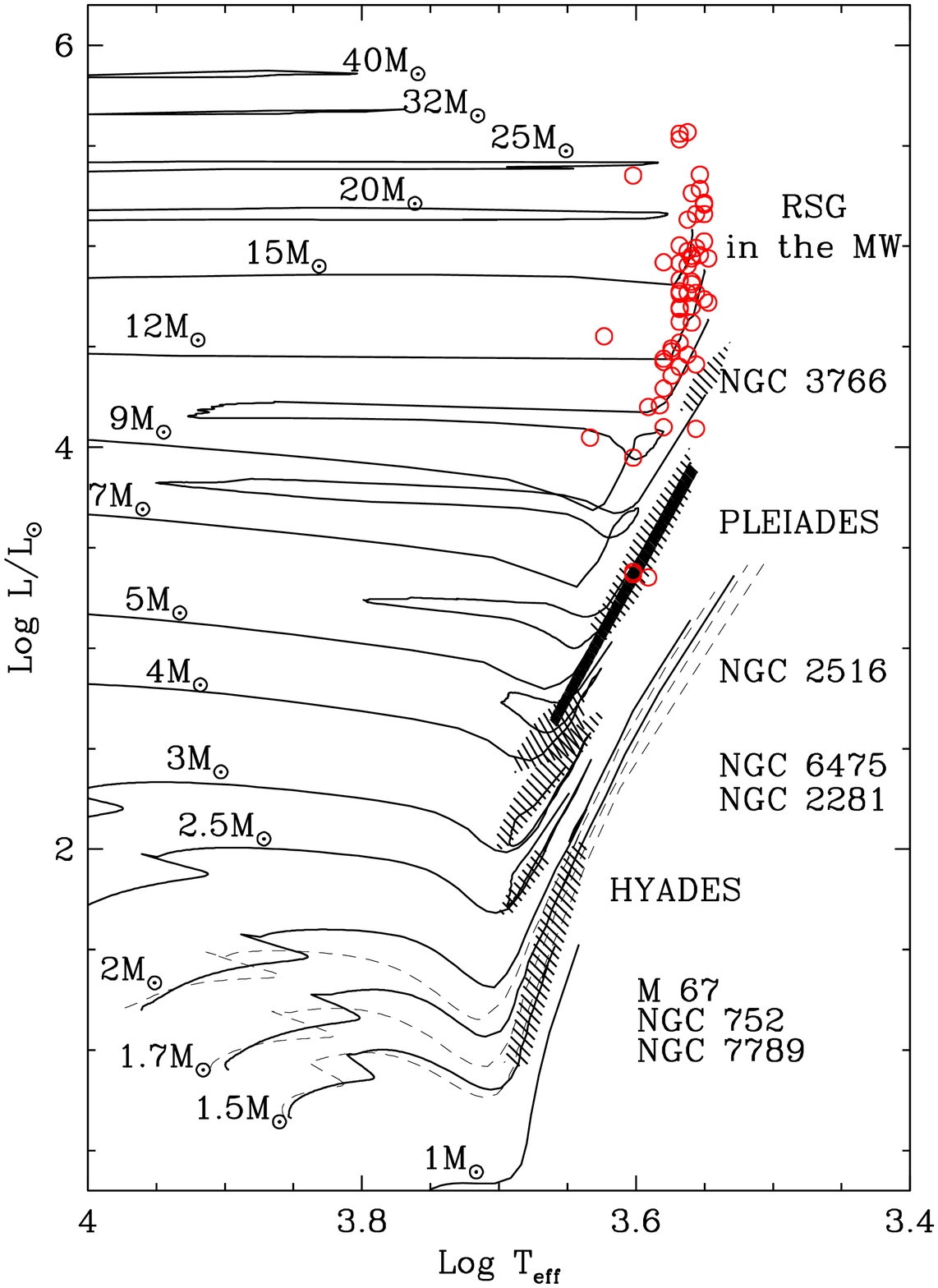}
\caption{{\it Left panel}: Profile of the CNO abundances through the stellar
envelope, as well as the profile of N/C and N/O ratios for the
rotating 25 M$_\odot$ when it reaches for the first time the RSG branch
(log(T$_{\rm eff}$) $\le$ 3.6). The x-axis is the Lagrangian mass coordinate,
and we show only the region above the convective core (7.80 M$_\odot$).
The light grey zones correspond to convective layers. The light
red zone corresponds to the region with 0.6 $< M_{\rm core}/M_{r} <$ 0.7,
corresponding to minimum values of the ratio of the core mass to
the total mass that is required to have a blue loop. Figure taken from Saio  {\em et al.} (\cite{2013MNRAS.433.1246S}).
{\it Right panel}: Evolutionary tracks for rotating models in the red part of the HR diagram. In the low mass range, a few non-rotating tracks (dashed lines) are shown. The shaded areas indicate
   the observations in clusters and associations 
as well as the position of galactic red supergiants obtained by Levesque {\em et al.} (\cite{2005ApJ...628..973L}). Figure taken from
Ekstr{\"o}m  {\em et al.} (\cite{2012A&A...537A.146E}).
}
\label{fig2}
\end{figure}

The pulsation properties of blue supergiants may be an interesting way to discriminate the blue supergiants which are evolving for the first time from the Main-Sequence to the red supergiant phase (hereafter called BSG1) from the blue supergiants which have already evolved through a red supergiant phase (hereafter called BSG2). Actually Saio {\em et al.}  (\cite{2013MNRAS.433.1246S}) have shown that BSG2 present many more excited pulsation modes than BSG1, when comparisons are made between stars of the same luminosity. 
The main reason for this is that BSG2 have lower actual masses since large amounts of mass have been lost by stellar winds in the
red supergiant phase. This increases the L/M ratio and thus allows many more pulsation modes to be excited. 
In order to validate this conclusion, we need to answer at least two questions: 1) are there observed blue supergiants showing
excited modes as predicted by models for BSG1 and BSG2? 2) Is there any other observational features indicating that BSG1
stars are actually on their first crossing of the HR gap and indicating that BSG2 have already evolved through a red supergiant phase?

To the first questions, we can answer yes.
There are indeed observed blue supergiants showing pulsations with frequencies predicted by the models for 
respectively the BSG1 and BSG2. Typically, the star HD 62150 presents pulsations compatible with those predicted by BSG1 models, while  a star like Rigel shows modes 
compatible with the predicted modes of BSG2 (see references and further examples in Saio  {\em et al.} \cite{2013MNRAS.433.1246S}). Thus, the modes predicted
by the models do appear quite in agreement with the observed pulsations in these cases. 

The answer of the second question first needs to find some criteria allowing to differentiate BSG1 from BSG2 other than the
pulsation properties. We can list the following characteristics:
\begin{itemize}
\item The surface composition is likely to be different between the BSG1 and BSG2. Indeed the BSG1 originating from initial masses
below about 40 M$_\odot$ undergo very little mass loss due to stellar winds during the Main-Sequence, thus their surface abundances can only be changed by internal mixing processes. For moderate rotation, one expects some increase of the
N/C ratio at the surface. Typically a solar metallicity 25 M$_\odot$ stellar models with a time-averaged velocity during the MS phase of 209 km s$^{-1}$   shows an increase of the N/C ratio with respect to the initial one by a factor 12 when it is a BSG1.
At the BSG2 stage, the increase is much higher (factor larger than 200) because the star has lost nearly half of its mass during the RSG phase and thus deeper and more processed material appear at the surface. 
\item Obviously, the actual mass of the star for a BSG1 star at a given luminosity will be larger than for a BSG2.
\item The  circumstellar environments of some BSG2 may still show some sign from the important mass loss episodes having occurred in the red supergiant phase. Of course depending on the time since the star has evolved away from the red supergiant phase, the relict of the red supergiant mass losses may have disappeared, at least in the vicinity of the star.
\end{itemize}
Let us now examine what these three criteria tell us about a star like Rigel whose initial mass is about 25 M$_\odot$. According to its pulsation properties, this star
should be a BSG2 (Saio  {\em et al.} \cite{2013MNRAS.433.1246S}). According to Przybilla  {\em et al.} (\cite{2006A&A...445.1099P}), the N/C ratio is around 12 times the initial one quite in agreement with the value predicted for a BSG1. Using spectroscopically determined gravity (log $g$=1.75) and estimating the radius (78.9 R$_\odot$) from interferometry and Hipparcos measurement, a mass of 13 M$_\odot$ is obtained (see references in Saio  {\em et al.} \cite{2013MNRAS.433.1246S}). Models indicate a mass between
10 and 13 M$_\odot$ for a 25 M$_\odot$ along the blue loop, supporting thus the idea that Rigel is a BSG2. Typical wind velocity for red supergiant winds is 10-15 km s$^{-1}$. If a few 100 thousand years separate the red supergiant phase from the blue supergiant one, then at least part of the matter ejected at the end of the red supergiant phase would be at about 6 10$^5$ stellar radii 
from the BSG2 (taking 79 R$_\odot$ for the radius of the BSG2).  Recent interferometric observations of Rigel (Kaufer {\em et al.} \cite{2012ASPC..464...35K})  show a complex circumstellar environment around Rigel at distance inferior to 1-2 solar radii. 
Therefore, the region investigated is much too near the star to probe mass loss episodes which might have occurred during a previous red supergiant phase. Thus this characteristic cannot be used
to infer the past history of Rigel.

To conclude, in case of Rigel, two criteria, the one using pulsation and the one concerning the actual mass of the star support the idea that this star is a BSG2, while
surface abundances favour a BSG1 solution.
At the moment, no obvious
solution can be proposed to level off the apparent contradiction between the abundance and mass and pulsation properties of Rigel.
If Rigel was initially a truly 25 M$_\odot$, how can the surface abundances not be changed after losing mass down to 13 M$_\odot$? 
In Fig.~\ref{fig2}, we can see that the loss of only 4 M$_\odot$ would be sufficient to increase significantly the N/C ratio well above the observed enhancement factor of 12.
At face, the observed feature for Rigel points towards less mixing in the radiative envelope, allowing thus to remove larger amount of mass without too much altering the surface abundances.
Of course, such tests should be applied to more stars in order to see whether some general trends appear. 

\section{Puzzle 3: How do the most massive stars going through a Red Supergiant phase end their life?}

The final state of star in the mass-range 15-25 M$_\odot$ is strongly dependent on the very uncertain mass-loss rates during the RSG phase, and can be as diverse as RSG, YSG, and LBV (Georgy \cite{2012A&A...538L...8G}, Groh {\em et al.} \cite{2013A&A...550L...7G}).

Type IIP supernovae are characterized by light curves showing a plateau for periods which lasts typically 100 days. This plateau comes from the fact that the pseudophotosphere formed at the recombination front of hydrogen remains more or less at the same position due to two counteracting effects: 1) as radiation cools the photosphere, the ionization front recedes inwards in mass
2) at the same time, due to expansion, mass moves outwards. This implies that the radius of the pseudophotosphere remains constant for a while and at a temperature corresponding to that of H recombination, i.e. around 6000 K. This produces the plateau in luminosity (see e.g. more details for instance in Kasen \& Woosley \cite{2009ApJ...703.2205K}). This kind of light curve occurs when a star with an extended H-rich envelope explodes and thus their progenitors are expected to be red supergiants. 

The most direct way to constrain the type IIP progenitors is to search for progenitors in archive images of the region where such supernovae have been observed. This has been the strategy by Smartt {\em et al.} (\cite{2009MNRAS.395.1409S}). They studied 20 type IIP events for which images before explosion was available. In five cases they indeed found a red supergiant progenitor. The other 15 cases are inconclusive because the events occurred in too crowded regions or only an upper limit for the luminosity of the progenitor was obtained. Using stellar tracks, they provided masses or upper mass limits for these 20 supernovae.
The lowest initial mass they found is 8+1-1.5 M$_\odot$ and the maximum initial mass is around 16.5+-1.5 M$_\odot$. We focus here on discussing the upper limit for the mass they found.
Note that Smartt {\em et al.} (\cite{2009MNRAS.395.1409S}) consider this upper mass limit as statistically significant at 2.4 $\sigma$ confidence.
If stars above, let us say 17-18 M$_\odot$ do not produce type IIP SNe, then of course the question is what other kinds of core collapse supernova do occur?

If we look at Fig.~\ref{fig2} (right panel), one sees first that red supergiants are observed well above the empirical upper limit for the progenitors of type IIP given by Smartt {\em et al.} (\cite{2009MNRAS.395.1409S}).
Second, one sees also 
that the tracks plotted have a different behavior for what concerns the red supergiant stage depending whether the initial mass of the star is below or above that empirical upper limit. 
Indeed, the tracks below that mass limit end their evolution as red supergiants, while stars with higher initial masses do not end their evolution as a red supergiant. Due to heavy mass loss, they evolve back in the blue part of the HR diagram. 
It happens that the 20 and 25 M$_\odot$ present at their presupernova stage a spectrum similar to Luminous Blue Variable stars (Groh {\em et al.} \cite{2013A&A...550L...7G}). Due to the low mass of hydrogen at their surface, they will probably not produce a type IIP SN event but more likely explode as a type IIb SN. Very interestingly, the present 20 M$_\odot$ ends its lifetime when its effective temperature is near one of the critical limits obtained by Vink {\em et al.} (\cite{1999A&A...350..181V}) where the stellar winds show rapid and strong changes (bistability limit\footnote{The concept of a bi-stability jump has been first described by Pauldrach \& Puls (\cite{1990A&A...237..409P}) }). It happens that this model, due to this effect, presents strong variations of its mass loss just before the explosion and this results in bumps in the radio light curves due to interactions between the ejecta and the matter released by these mass loss episode (Moriya et al. \cite{2013arXiv1306.0605M}). Interestingly such bumps in radio light curves are indeed observed in the case for instance of SN IIb 2001ig (Ryder v \cite{2004MNRAS.349.1093R}). This gives support to the present models and indicates that single star evolutionary scenarios exist for producing type IIb supernovae.

\begin{figure}
\includegraphics[width=2.0in,height=2.0in]{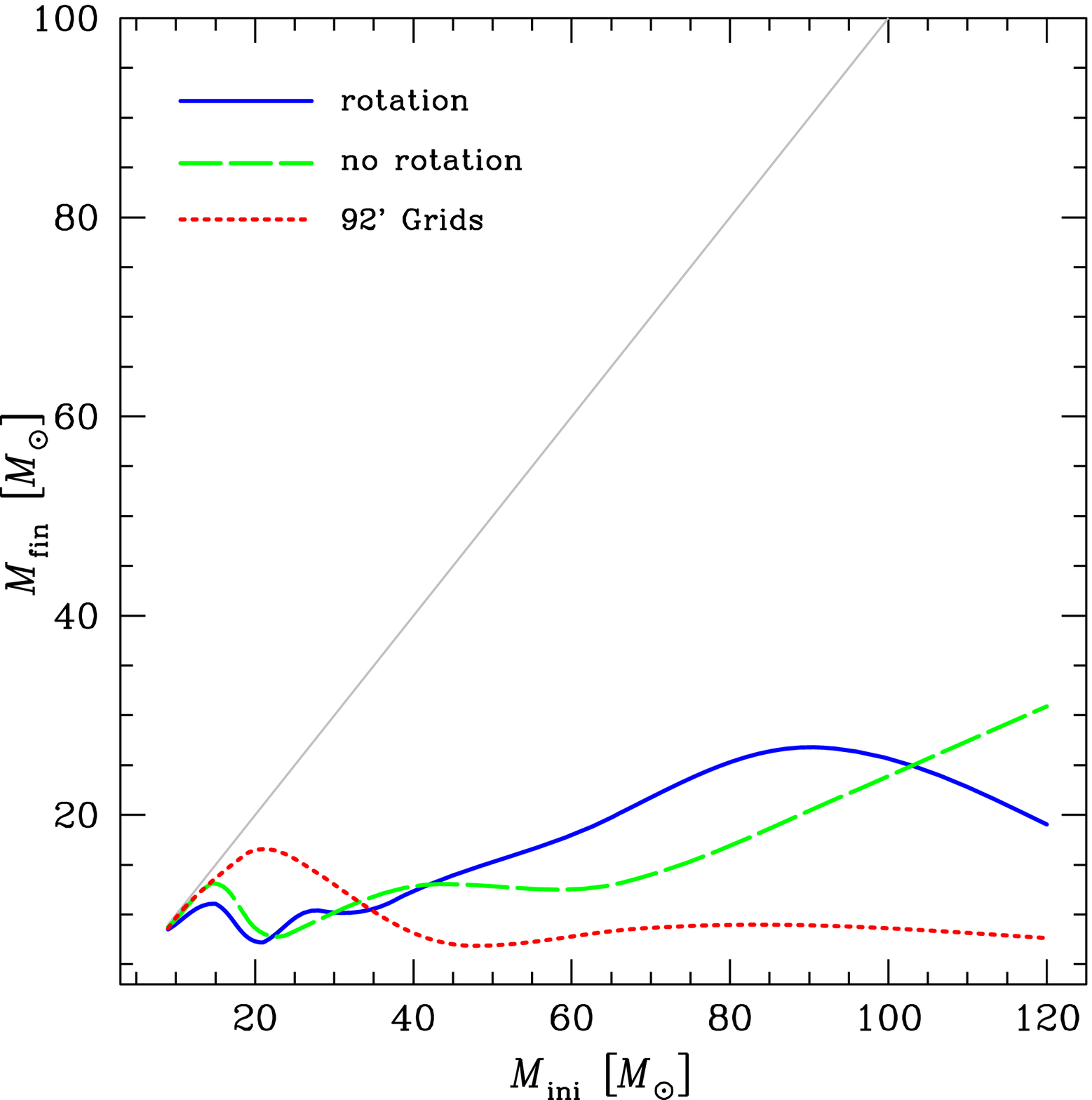}
\hspace{1cm}
\includegraphics[width=2.1in,height=2.1in]{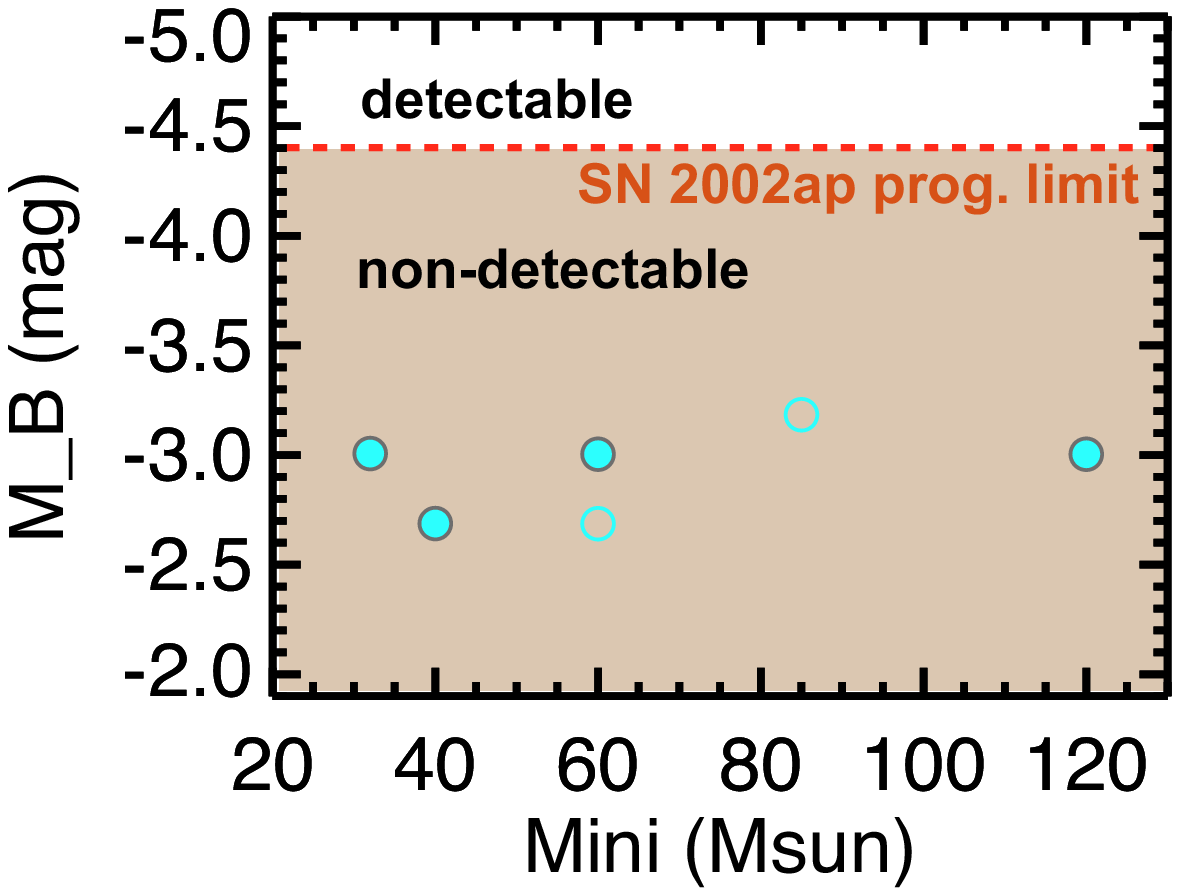}
\caption{{\it Left panel}: Final mass vs initial mass. Comparison between the non-rotating models (dashed green line) and rotating models (solid blue line) of Ekstr{\"o}m  {\em et al.} (\cite{2012A&A...537A.146E}), and the normal rates grids models of Schaller   {\em et al.} (\cite{1992A&AS...96..269S}, dotted red line). In grey, the line of constant mass. Figure taken from Ekstr{\"o}m  {\em et al.} (\cite{2012A&A...537A.146E}).
{\it Right panel}: Absolute B magnitude of our models that are SN Ic progenitors
(open and filled cyan circles) compared to the present day lowest observed upper magnitude
limit for a progenitor of a SN Ic (colored horizontal dashed lines followed by
the shortened SN label). Open
(filled) symbols correspond to non-rotating (rotating) models.  
Note that all models lie below the detectability limit of all SN Ic. Figure taken from Groh {\em et al.} (\cite{2013arXiv1308.4681G}).
}
\label{fig3}
\end{figure}

\section{Puzzle 4: What is the nature of the SN Ibc progenitors?}

Type Ibc supernovae are produced by the collapse of stars having lost their H-rich envelope. This loss may have occurred through stellar winds and/or through a Roche Lobe Overflow (RLOF) event in close binaries. One can wonder whether it is important to know how this mass loss occurred, {\it i.e.} through stellar winds or a RLOF event? This has some importance indeed, since these two mechanisms will
make different predictions about the mass of the core at the time of the SN explosion, they will also linked these types of SNe events to different initial mass ranges and thus make different predictions concerning their frequency and the dependance of their frequency
on the metallicity.

Twelve Ibc supernovae with deep pre-explosion images are discussed in Smartt (\cite{2009ARA&A..47...63S} and see further references therein) and in
Eldridge {\em et al.}  (\cite{2013arXiv1301.1975E}).
In none of these cases, a progenitor has been detected.
Comparing the upper limits for the progenitor's luminosity (in right panel of Fig.~\ref{fig3}, the lowest of this upper limit is indicated) with the observed luminosity of WR stars suggests at 90\% confidence level that the hypothesis according to which the WR stars
we observe in the local group are the only progenitors of type Ibc supernovae is false (Smartt \cite{2009ARA&A..47...63S}). 

However, the observed WN and WC stars may not be at their pre-supernova stage. Even if time is short before the explosion, still some changes of the effective temperature and  luminosity may make them much less easy to detect. In Groh  {\em et al.} (\cite{2013arXiv1308.4681G}), the spectrum of the presupernova models at solar metallicity given by Ekstr{\"o}m et al (\cite{2012A&A...537A.146E}) have been computed using the wind-atmosphere code CFMGEN (Hillier \& Miller \cite{1998ApJ...496..407H}). It happens that at the presupernova stage, the stars progenitors of type Ic supernovae are very hot, and have a typical spectrum of a WO star. They would have luminosities in the different bands well below
the upper limit determined by Smartt (\cite{2009ARA&A..47...63S}). This goes exactly in the same direction as the conclusion by Yoon {\em et al.} (\cite{2012A&A...544L..11Y})
and can be a reasonable explanation for the non-detection mentioned above. Thus the non-detection mentioned above cannot be taken at the moment as an argument for invoking other stars than the observed WR stars
as progenitors for this type of supernovae.
It will be interesting to push down the dectability limit and to see whether a star will appear with a luminosity predicted by 
the models of single stars or not. It seems that such a star may have been detected in the case of a SN Ib (Cao  {\em et al.} \cite{2013arXiv1307.1470C}; see also discussion in Groh  {\em et al.}  \cite{2013arXiv1307.8434G}).


\section{Conclusion}
In case positive answers would be given to the four questions below,
1.- Do slow, non-evolved rotators which are N-rich have a surface magnetic field and rotate differentially in their interior?
2.- Do most of Alpha-Cygni variables show strong N-enrichments?
3.- Do most stars with masses between 18 and 30 M$_\odot$ end their lifetimes as blue supergiants and/or explode as type IIL, or IIb supernova?
4.- Are the progenitors of type Ic Sne, WO type stars with properties characteristic of single star evolution?
then this would provide some support to the present models. If negative answers are obtained, we can remind ourselves this sentence of Jean Rostand:
{\it One must credit an hypothesis with all that has been discovered 
in order to demolish it.} 

\end{document}